\documentclass[aip,rsi,amsmath,amssymb,reprint,]{revtex4-1}

\usepackage{graphicx}
\usepackage{dcolumn}
\usepackage{bm}
\usepackage{booktabs}
\usepackage{multirow}

\begin{document}

\title{High detection efficiency silicon single-photon detector with a monolithic integrated circuit of active quenching and active reset}

\author{Yu-Qiang Fang}
 \affiliation{Hefei National Laboratory for Physical Sciences at the Microscale and Department of Modern Physics, University of Science and Technology of China, Hefei 230026, China}
 \affiliation{CAS Center for Excellence in Quantum Information and Quantum Physics, University of Science and Technology of China, Hefei 230026, China}
\author{Kai Luo}
 \affiliation{Sichuan Institute of Solid-State Circuits, China Electronics Technology Group Corp., Chongqing 400060, China}
 \affiliation{College of Optoelectronic Engineering, Chongqing University, Chongqing 400044, China}
\author{Xing-Guo Gao}
 \affiliation{Sichuan Institute of Solid-State Circuits, China Electronics Technology Group Corp., Chongqing 400060, China}
\author{Gai-Qing Huo}
 \affiliation{Sichuan Institute of Solid-State Circuits, China Electronics Technology Group Corp., Chongqing 400060, China}
\author{Ang Zhong}
 \affiliation{Sichuan Institute of Solid-State Circuits, China Electronics Technology Group Corp., Chongqing 400060, China}
\author{Peng-Fei Liao}
 \affiliation{Sichuan Institute of Solid-State Circuits, China Electronics Technology Group Corp., Chongqing 400060, China}
\author{Pu Pu}
 \affiliation{Sichuan Institute of Solid-State Circuits, China Electronics Technology Group Corp., Chongqing 400060, China}
\author{Xiao-Hui Bao}
 \affiliation{Hefei National Laboratory for Physical Sciences at the Microscale and Department of Modern Physics, University of Science and Technology of China, Hefei 230026, China}
 \affiliation{CAS Center for Excellence in Quantum Information and Quantum Physics, University of Science and Technology of China, Hefei 230026, China}
\author{Yu-Ao Chen}
 \affiliation{Hefei National Laboratory for Physical Sciences at the Microscale and Department of Modern Physics, University of Science and Technology of China, Hefei 230026, China}
 \affiliation{CAS Center for Excellence in Quantum Information and Quantum Physics, University of Science and Technology of China, Hefei 230026, China}
\author{Jun Zhang}
 \email{zhangjun@ustc.edu.cn}
 \affiliation{Hefei National Laboratory for Physical Sciences at the Microscale and Department of Modern Physics, University of Science and Technology of China, Hefei 230026, China}
 \affiliation{CAS Center for Excellence in Quantum Information and Quantum Physics, University of Science and Technology of China, Hefei 230026, China}
\author{Jian-Wei Pan}
 \affiliation{Hefei National Laboratory for Physical Sciences at the Microscale and Department of Modern Physics, University of Science and Technology of China, Hefei 230026, China}
 \affiliation{CAS Center for Excellence in Quantum Information and Quantum Physics, University of Science and Technology of China, Hefei 230026, China}

\date{\today}

\begin{abstract}
Silicon single-photon detectors (SPDs) are key devices for detecting single photons in the visible wavelength range.
Photon detection efficiency (PDE) is one of the most important parameters of silicon SPDs, and increasing PDE is highly required for many applications.
Here, we present a practical approach to increase PDE of silicon SPD with a monolithic integrated circuit of active quenching and active reset (AQAR).
The AQAR integrated circuit is specifically designed for thick silicon single-photon avalanche diode (SPAD) with high breakdown voltage (250-450 V), and then fabricated via the process of high-voltage 0.35-$\mu$m bipolar-CMOS-DMOS.
The AQAR integrated circuit implements the maximum transition voltage of $\sim$ 68 V with 30 ns quenching time and 10 ns reset time, which can easily boost PDE to the upper limit by regulating the excess bias up to a high enough level.
By using the AQAR integrated circuit, we design and characterize two SPDs with the SPADs disassembled from commercial products of single-photon counting modules (SPCMs).
Compared with the original SPCMs, the PDE values are increased from 68.3\% to 73.7\% and 69.5\% to 75.1\% at 785 nm, respectively, with moderate increases of dark count rate and afterpulse probability.
Our approach can effectively improve the performance of the practical applications requiring silicon SPDs.
\end{abstract}

\maketitle

\section{Introduction}

\begin{figure*}[tbp]
\centering
\includegraphics[width=0.65\linewidth]{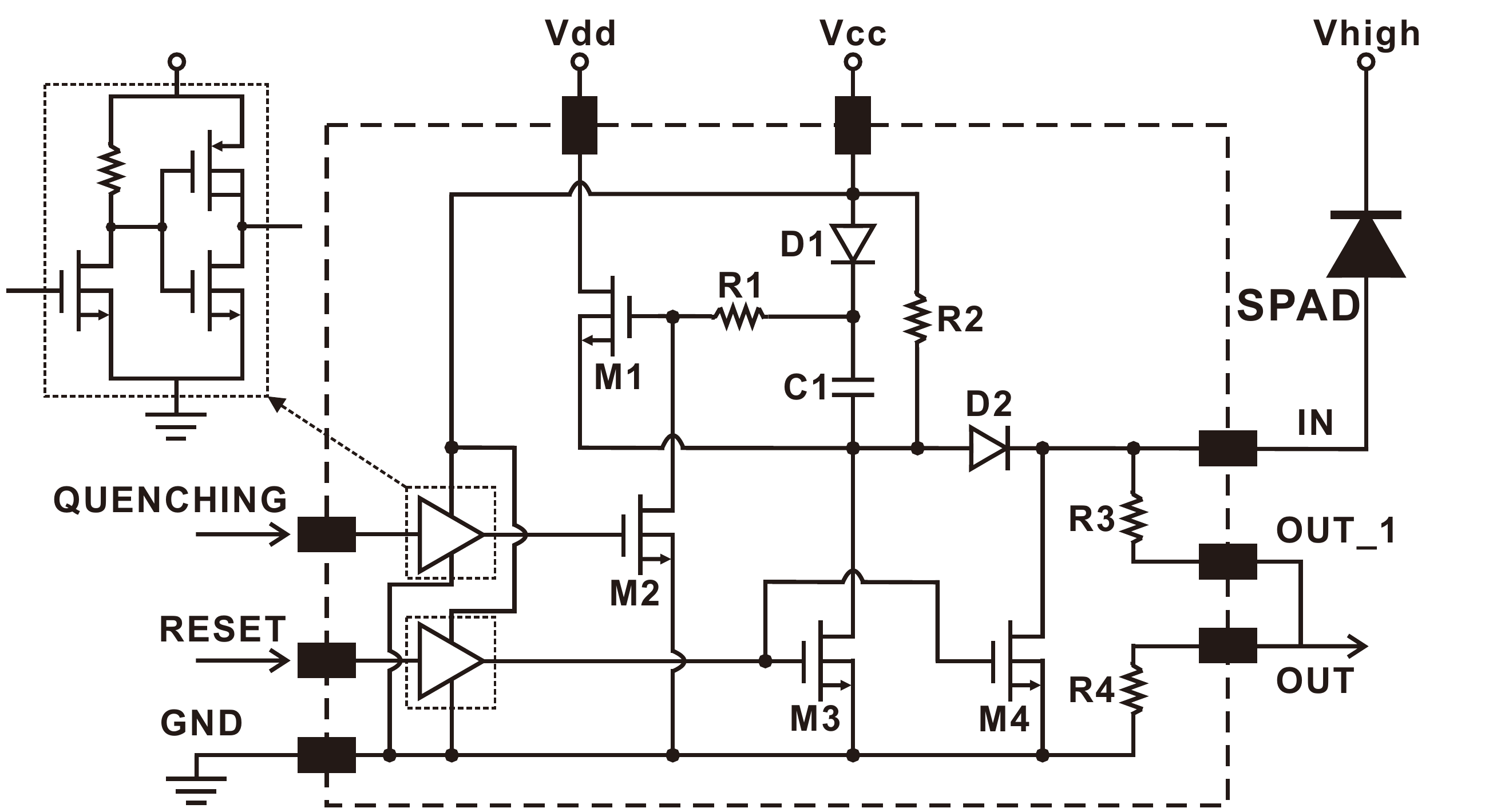}
\caption{Schematic diagram of the AQAR circuit.}
\label{fig1}
\end{figure*}

Single-photon detectors (SPDs) are the most sensitive tools to detect extremely weak light and are widely used in numerous applications~\cite{SPD11}.
Currently, there are several SPD technologies such as photomultiplier tubes (PMTs), superconducting nanowire single-photon detectors~\cite{SNSPD13,SNSPD17}, and single-photon avalanche diodes (SPADs)~\cite{Cova04,Zhang15},
among which using SPADs is the primary solution for practical applications due to the advantages of small size and low cost.
In the visible wavelength range, silicon SPADs are normally used to detect single photons~\cite{DDD93}.
The key parameters for characterization include photon detection efficiency (PDE), dark count rate (DCR), afterpulse probability ($P_{ap}$), maximum count rate ($CR_{max}$), and timing jitter ($T_{jitter}$).

A silicon SPD is composed of a SPAD and the corresponding quenching circuit~\cite{Cova96}.
There are three kinds of silicon SPADs~\cite{Cova04}.
The first one is the thin silicon SPAD, e.g., PDM photon counting module~\cite{PDM}, which is a planar epitaxial device with thin depletion layer of a few micrometers, low breakdown voltage ($V_{br}$) (15-40 V), and small active area.
Such SPAD exhibits $\sim$ 50\% PDE at 550 nm, $\sim$ 50 cps DCR, and $\sim$ 50 ps $T_{jitter}$.
The second one is the thick silicon SPAD, e.g., single-photon counting module (SPCM)~\cite{SPCM}, which is a reach-through device with thick depletion layer of tens of micrometers, high $V_{br}$ (250-450 V), and large active area.
SPCM exhibits $\sim$ 70\% PDE at 650 nm, 25 cps DCR (best grade), and $\sim$ 400 ps $T_{jitter}$.
The third one is the SPAD device fabricated via the complementary metal–oxide–semiconductor (CMOS) technology, which exhibits $\sim$ 45\% PDE at 450 nm~\cite{TTZ07}.
The quenching circuit is crucial for SPAD to achieve high performance, in which SPAD is operated either in gating mode or in free-running mode. So far,
high detection efficiency silicon SPDs using high-frequency gating techniques have been reported~\cite{TYD10,SNT14,Zhou17}.
For the free-running mode, passive quenching is a simple solution~\cite{Cova96,Cova04,HLZ10,GRG10},
however, it suffers from low $CR_{max}$ and slow recovery.
Active quenching overcomes the drawbacks of passive quenching while brings the challenges to the electronics design
in terms of short quenching time and reset time.
So far, various approaches to active quenching circuits have been reported~\cite{GCZ96,ZGC00,ZGG02,ZLG03,TGZ08,Mario09,ARG16,Mario17,ALR17,AGR18,ACG18}.
For instance, the circuits of SPCMs adopt an approach of double quenching, i.e., passive quenching at the cathode and active quenching at the anode~\cite{patent}.
The circuits use discrete devices and the excess bias voltage is limited due to the breakdown voltage of the transistors.

In this paper, we present an approach to increase PDE of silicon SPD to the upper limit with a monolithic integrated circuit of active quenching and active reset (AQAR). The AQAR integrated circuit is specifically designed for the thick silicon SPAD and is fabricated via the process of high-voltage 0.35-$\mu$m bipolar-CMOS-DMOS (BCD),
with the maximum transition voltage of $\sim$ 68 V, 30 ns quenching time, and 10 ns reset time.
We also develop two silicon SPDs using the AQAR chips and SPADs disassembled from commercial products of SPCMs.
The characterization results show that compared with the original SPCMs, the PDE values of SPDs can be increased from 68.3\% to 73.7\% and 69.5\% to 75.1\% at 785 nm, respectively, with moderate increases of DCR and $P_{ap}$.

\section{Integrated circuit design and fabrication}

Fig.~\ref{fig1} illustrates the schematic diagram of the AQAR integrated circuit.
The anode of the silicon SPAD is connected to the IN pin of the chip.
In order to achieve fast voltage transition, four n-channel DMOS transistors (M1-M4) are designed, among which M1 and M2 are used for quenching switches while M3 and M4 are used for reset switches.
The input signals from the QUENCHING pin and the RESET pin are in the transistor-transistor logic (TTL) level.
They are converted into signals with 12 V amplitude ($V_{cc}$) to improve the drive capability by two drivers that are integrated into the AQAR circuit,
as shown in the inset of Fig.~\ref{fig1}.

In a quiescent condition, the quenching signal is at the logic level HIGH to make M2 on while the reset signal is at the logic level LOW to make M3 and M4 off.
Since M2 turns on, the gate voltage of M1 is below the threshold and thus M1 turns off.
The voltage at the IN node is close to zero. The bias voltage ($V_{high}$) is above $V_{br}$. The SPAD is ready for single-photon detection, and the whole circuit is in the quiescent state.
Once an avalanche occurs, the avalanche current immediately produces a voltage drop on the resistors of R3 and R4.
R3 is a relatively high-value resistor to quench the avalanche passively and R4 is used as a voltage divider to produce an output signal for discrimination.
The inverse output signal from the discriminator is connected with the QUENCHING pin.
Since the QUENCHING signal is at the logic level LOW, M2 turns off and M1 turns on.
The bootstrap capacitor C1 maintains M1 in the linear region of operation so that the voltage at the IN node keeps growing to $V_{dd}$.
In such a way, the voltage difference between the cathode and anode of the silicon SPAD is changed to $V_{high}-V_{dd}$, which is below $V_{br}$.
Therefore, the avalanche process is completely quenched.
After a controlled hold-off time, the signal at the RESET pin is set at the logic level HIGH, which results in that
M3 and M4 turn on promptly to reset the voltage at the IN node to zero.
Then, the AQAR circuit is recovered to the initial quiescent state for the next single-photon detection.

\begin{figure*}[tbp]
\centering
\includegraphics[width=0.7\linewidth]{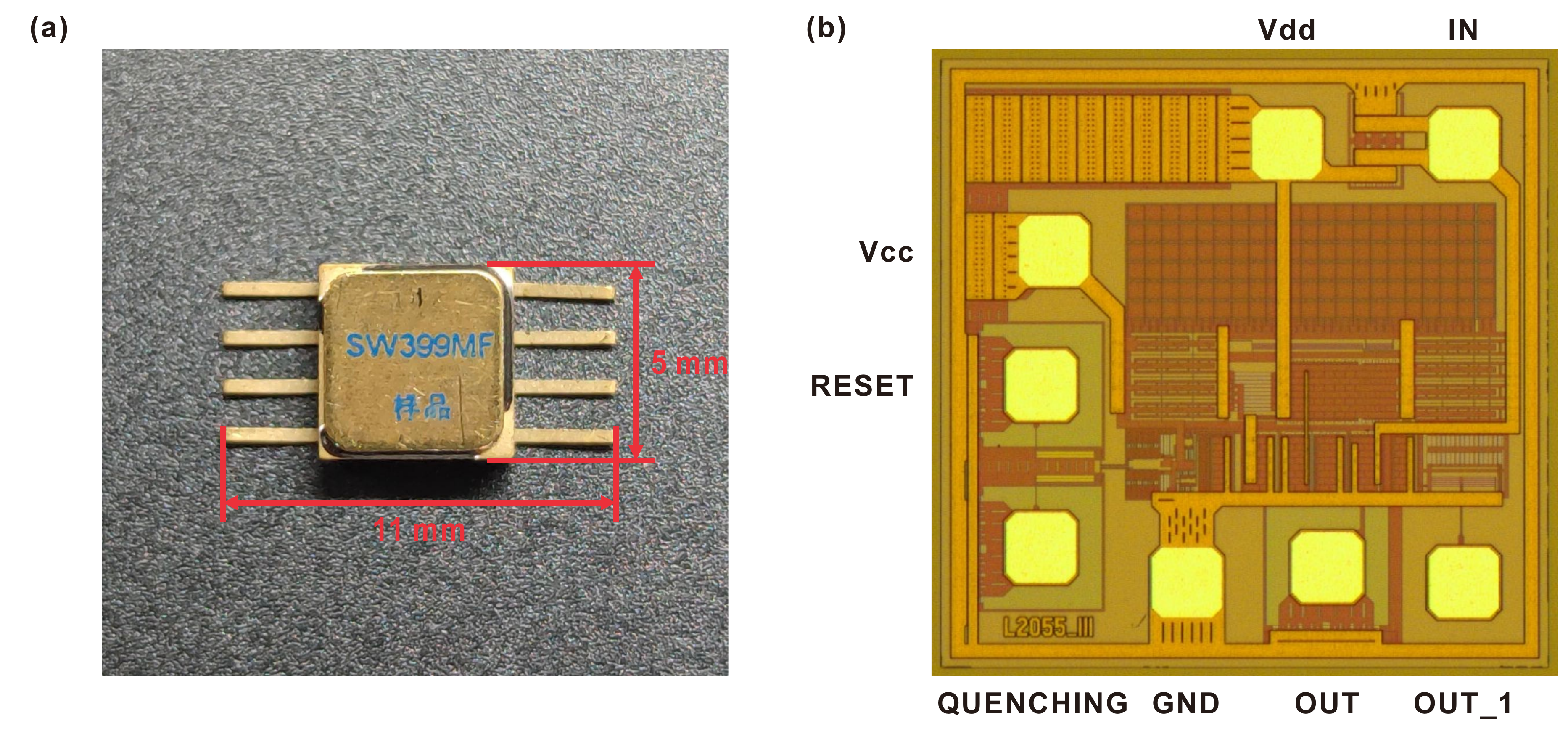}
\caption{(a) Photo of the AQAR chip with an overall package dimension of 11 mm $\times$ 5 mm. (b) Microphotograph of the die inside the AQAR chip.}
\label{fig2}
\end{figure*}

The AQAR chip is fabricated via the process of high-voltage 0.35-$\mu$m BCD.
The BCD process combines the advantages of three different processing technologies into a single chip, among which DMOS is specifically suited for high-voltage circuits.
During the process, the triple well process with a deep n-well is used to fabricate the isolated thick gate oxide NLDMOS (N-channel Laterally Diffused Metal Oxide Semiconductor) transistors and high-voltage diodes.
The drain-to-source, source-to-substrate, and drain-to-substrate breakdown voltages of the transistors are $\sim$ 100 V.
The reverse breakdown voltages of the diodes are $\sim$ 80 V.
The resistors are implemented by injecting two layers of polysilicon among four layers of metal wiring, and
the resistance values are regulated by the doping concentration of polysilicon.
A metal capacitor filled with a nitride medium is also integrated.
The breakdown voltage between the plates of the capacitor is 14 V, and the breakdown voltages between the plates and the substrate are $\sim$ 100 V.
The AQAR chip is then sealed inside a ceramic small outline package with an overall dimension of 11 mm $\times$ 5 mm, as shown in Fig.~\ref{fig2}a.
Fig.~\ref{fig2}b shows the microphotograph of the die inside the AQAR chip with a
dimension of 1.1 mm $\times$ 1.1 mm.
The dimension of the pad for bonding is 130 $\mu$m $\times$ 130 $\mu$m.

\section{SPD module and characterization}

\begin{figure*}[tbp]
\centering
\includegraphics[width=0.7\linewidth]{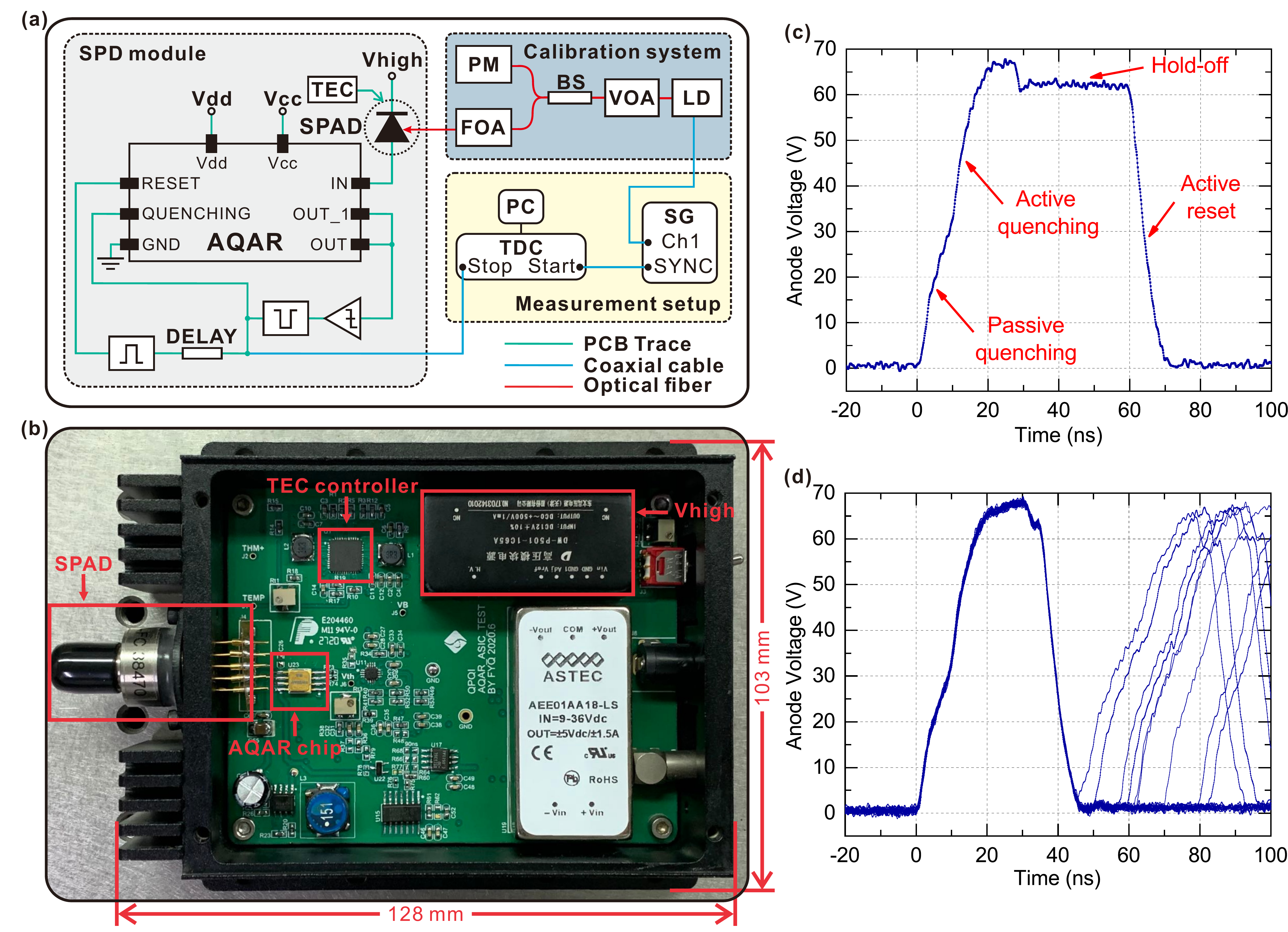}
\caption{(a) Schematic diagram of the SPD module and the experimental setup for single-photon characterization at 785 nm. (b) The photo of the SPD module. (c) A typical voltage waveform captured by an oscilloscope at the anode of the silicon SPAD. (d) The voltage waveform persistence at the anode of the silicon SPAD with $V_{dd}$ = 68 V and the minimal hold-off time shows that the intrinsic dead time is $\sim$ 42 ns.}
\label{fig3}
\end{figure*}

\begin{figure*}[tbp]
\centering
\includegraphics[width=0.7\linewidth]{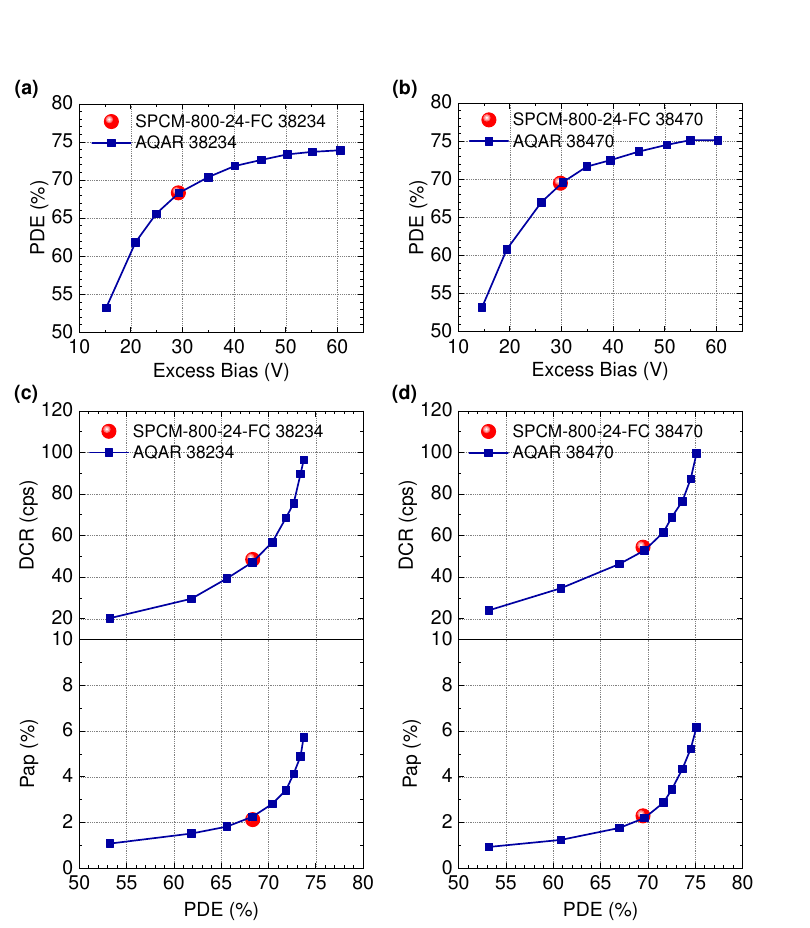}
\caption{PDE versus excess bias of AQAR 38234 (a) and AQAR 38470 (b). DCR and afterpulse probability versus PDE of AQAR 38234 (c) and AQAR 38470 (d). Red circles represent the results of SPCMs.}
\label{fig4}
\end{figure*}

Further, we design an SPD module with the AQAR chip. Fig.~\ref{fig3}a illustrates
the schematic diagram of the SPD module, the corresponding single-photon calibration system, and the measurement setup.
Fig.~\ref{fig3}b exhibits the photo of the designed SPD module, with
a dimension of 128 mm $\times$ 103 mm $\times$ 33 mm, a weight of 380 g, and a power dissipation of $\sim$ 3.5 W.
In the module, the silicon SPAD is biased with a tunable high-voltage power supply ($V_{high}$).
The temperature of the SPAD is regulated by a thermoelectric cooler (TEC) controller.
The anode of the SPAD is connected as close as possible to the IN pin of the AQAR chip.
The avalanche signal from the OUT pin is discriminated by a high-speed comparator,
and the subsequent monostable circuit produces a 20 ns LOW-level effective signal connected with the QUENCHING pin.
The propagation delay between the OUT pin and the QUENCHING pin is calibrated to be as low as 3 ns.
Also, the monostable circuit produces a 10 ns HIGH-level effective signal connected with the RESET pin after a certain hold-off time. The hold-off time can be tuned from 0 to 180 ns by delay lines, which is used to reduce the afterpulse probability.
Fig.~\ref{fig3}c plots a typical voltage waveform captured by an oscilloscope at the IN pin, from which one can observe that
the overall dead time of SPD is contributed by four processes, i.e., passive quenching, active quenching, hold-off duration, and active reset. By setting $V_{dd}$ to the maximum value, i.e., $\sim$ 68 V, the quenching time and the reset time are measured to be 30 ns and 10 ns, respectively. Moreover, by setting the hold-off time to the minimum value, the voltage waveform persistence, as shown in Fig.~\ref{fig3}d, indicates that the intrinsic dead time of SPD is $\sim$ 42 ns.

Following the single-photon calibration scheme~\cite{Zhang15}, one channel (CH1) of the signal generator (SG) generates a clock of 100 kHz to drive a picosecond pulsed laser diode (LD, PicoQuant) at 785 nm.
The laser pulses with a full width at half maximum (FWHM) below 70 ps pass through a variable optical attenuator (VOA), a 99:1 beam splitter (BS), and a fixed optical attenuator (FOA). Then,
the intensity of the laser pulses is attenuated down to a level of a mean photon number per pulse of 1.
The 99\% port of the BS is monitored with a power meter (PM, Newport). A time-to-digital converter (TDC, PicoQuant) is used to perform measurements and transmit data to a personal computer (PC), so that the parameters of PDE, DCR, $P_{ap}$, $CR_{max}$, and $T_{jitter}$ can be calculated.

\begin{table*}[htbp]
\centering
\caption{Parameter comparison between SPCMs and SPDs with the AQAR integrated circuits for the same SPADs at 785 nm. $\Delta_{PDE}$ represents the relative increase value of PDE.}
\label{table1}
\begin{tabular}{ccccccc}
\hline
Part number & PDE & DCR (cps) & $P_{ap}$ & $T_{jitter}$ (ps) & $CR_{max}$ (MHz) & $\Delta_{PDE}$\\
\hline
SPCM 800 24 FC 38234 & 0.683 & 48.5 & 0.021 & 430 & 31.0 & \multirow{2}*{0.08} \\
AQAR 38234 & 0.737 & 96.4 & 0.057 & 370 & 10.1 & ~ \\
\hline
SPCM 800 24 FC 38470 & 0.695 & 54.4 & 0.023 & 460 & 31.0 & \multirow{2}*{0.08} \\
AQAR 38470 & 0.751 & 99.7 & 0.062 & 390 & 10.2 & ~ \\
\hline
\end{tabular}
\end{table*}

\section{Results and discussion}

We perform single-photon characterization of SPDs at 785 nm with two silicon SPADs disassembled from commercial products of SPCMs (part numbers: SPCM-800-24-FC 38234 and SPCM-800-24-FC 38470) and compare the parameters between our SPDs (part numbers: AQAR 38234 and AQAR 38470) and the original SPCMs.
The operating temperatures of SPADs in our SPDs are regulated as that in SPCMs, i.e., 252 K, by measuring the resistance of the thermistor in the SPADs.
For our SPDs, the hold-off time is set to 40 ns and the overall dead time is $\sim$ 82 ns.
Fig.~\ref{fig4} shows the measured results in two cases.
For SPCMs, $V_{br}$ of two SPADs is calibrated to be 339.5 V and 355.0 V at 252 K, and PDE is measured to be 68.3\% and 69.5\%, respectively, with $\sim$ 30 V $V_{ex}$.

Fig.~\ref{fig4}a and Fig.~\ref{fig4}b plot PDE as a function of excess bias.
With the increase of excess bias, PDE continues to increase. When $V_{ex}$ reaches $\sim$ 55 V, PDE becomes flat, which implies that PDE is very close to the upper limit.
The parameters of DCR and $P_{ap}$ as a function of PDE for two SPDs are also plotted in Fig.~\ref{fig4}c and Fig.~\ref{fig4}d, respectively, from which one can clearly observe that the PDE of our SPDs is significantly higher than SPCMs with a cost of moderate increases of DCR and $P_{ap}$. The detailed results for the comparison are listed in Table~\ref{table1}.
Compared with the original SPCMs, the PDE values of our SPDs are increased from 68.3\% to 73.7\% and 69.5\% to 75.1\%, respectively.
The relative increases ($\Delta_{PDE}$) for both SPDs reach $\sim$ 8\%.

\section{Conclusion}

In conclusion, we have reported an effective and practical approach to increase the detection efficiency of silicon SPD using a monolithic integrated circuit of AQAR that is dedicated to thick silicon SPAD with high breakdown voltage.
By applying the AQAR integrated circuit, the maximum transition voltage can reach $\sim$ 68 V with 30 ns quenching time and 10 ns reset time. Such an integrated circuit can easily push PDE to the upper limit.
We have designed SPDs with the integrated circuits and two SPADs disassembled from commercial SPCMs. Parameter comparison between our SPDs and the original SPCMs has been performed, which shows that PDE can be relatively increased by 8\%.
Our work can effectively improve the performance in diverse applications requiring high detection efficiency silicon SPDs.

\section*{acknowledgements}

This work has been supported by the National Key R\&D Program of China under Grant No.~2017YFA0304004, the National Natural Science Foundation of China under Grant No.~11674307, the Chinese Academy of Sciences, and the Anhui Initiative in Quantum Information Technologies.

\section*{Data Availability Statement}
The data that support the findings of this study are available from the corresponding author upon reasonable request.

\bibliography{AQAR}

\end{document}